\documentclass[11pt]{cernrep}
\usepackage{graphicx}
\usepackage{cite,./mcite}
\usepackage{graphicx}
\usepackage{psfrag}

\newcommand{\kt}{k_{T}}
\newcommand{\pt}{p_{T}}
\newcommand{\ptpair}{p_{T c\bar{c}}}

\newcommand{\VC}[1]{%
  \begin{tabular}[c]{l}%
    #1%
  \end{tabular}
}
\begin{document}

\title{Need for fully unintegrated parton densities}
\author{John Collins$^1$, Hannes Jung$^2$}
\institute{        $^1$ Physics Department,
        Penn State University,
        104 Davey Laboratory,
        University Park PA 16802,
        U.S.A.\\
          $^2$ DESY,
        Notkestra{\ss}e 85,
        22603 Hamburg,
        FRG}

\maketitle
\begin{abstract}
  Associated with the use of conventional integrated parton densities
  are kinematic approximations on parton momenta which result in
  unphysical differential distributions for final-state particles.  We
  argue that it is important to reformulate perturbative QCD results
  in terms of fully unintegrated parton densities, differential in all
  components of the parton momentum.
\end{abstract}

\section{Introduction}

Conventional parton densities are defined in terms of an integral over
all transverse momentum and virtuality for a parton that initiates a
hard scattering.  While such a definition of an integrated parton
density is appropriate for very inclusive quantities, such as the
ordinary structure functions $F_1$ and $F_2$ in DIS, the definition
becomes increasingly unsuitable as one studies less inclusive cross
sections.  Associated with the use of integrated parton densities are
approximations on parton kinematics that can readily lead to
unphysical cross sections when enough details of the final state are
investigated.  

We propose that it is important to the future use of pQCD that a
systematic program be undertaken to reformulate factorization results
in terms of fully unintegrated densities, which are differential in
both transverse momentum and virtuality.  These densities are called
``doubly unintegrated parton densities'' by Watt, Martin and Ryskin
\cite{MRW1,MRW2}, and ``parton correlation functions'' by
Collins and Zu \cite{CZ}; these authors have presented the reasoning
for the inadequacy, in different contexts, of the more conventional
approach.  The new methods have their motivation in contexts such as
Monte-Carlo event generators where final-state kinematics are studied
in detail.  Even so, a systematic reformulation for other processes to
use unintegrated densities would present a unified methodology. 

These methods form an extension of $\kt$-factorization, which has so
far been applied in small-$x$ processes and, as the CSS formalism, in
the transverse-momentum distribution of the Drell-Yan and related
processes.

\section{Inadequacy of conventional pdfs}

\begin{figure}
  \centering
  \includegraphics[scale=0.6]{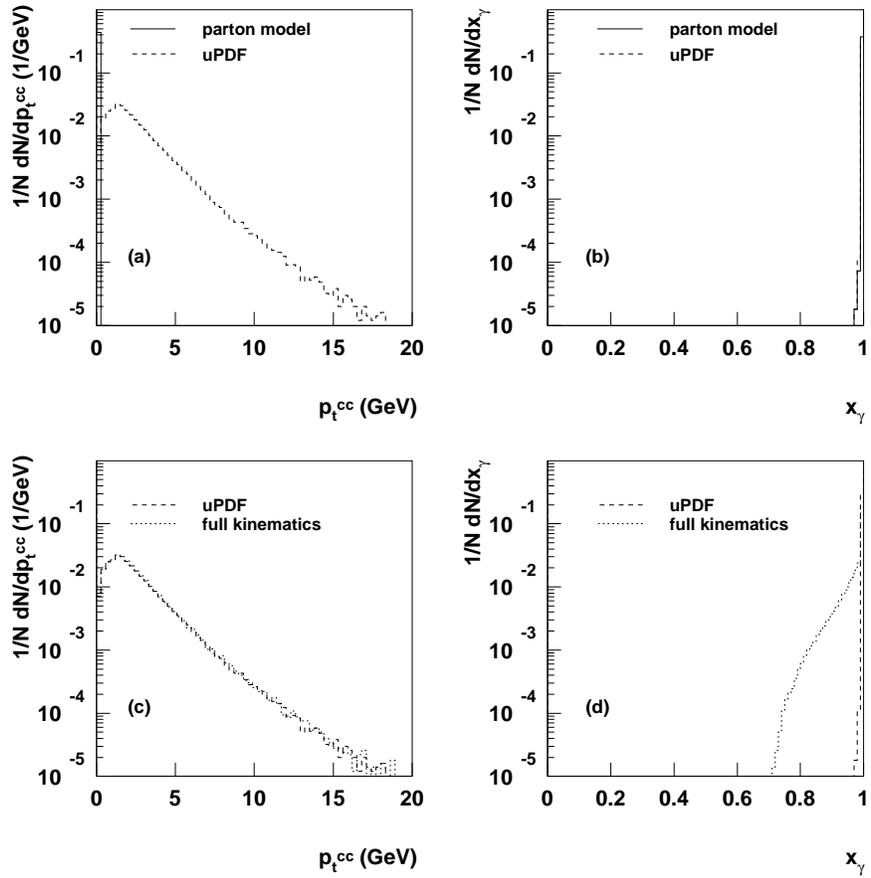}
  \caption{
    (a) and (b): Comparison between use of simple LO parton model
    approximation and of the use of $\kt$ densities for the $\pt$ of
    $c\bar{c}$ pairs in photoproduction, and for the $x_\gamma$.  (c) and
    (d): Comparison of use of $\kt$ densities and full simulation.  }
  \label{fig:ccbar.1}
\end{figure}

\begin{figure}
  \centering
  $
  \psfrag{q}{\small$q$}
  \psfrag{k}{\small$k$}
  \psfrag{p1}{\small$p_1$}
  \psfrag{p2}{\small$p_2$}
  \psfrag{P}{\small$P$}
  \VC{\includegraphics[scale=0.5]{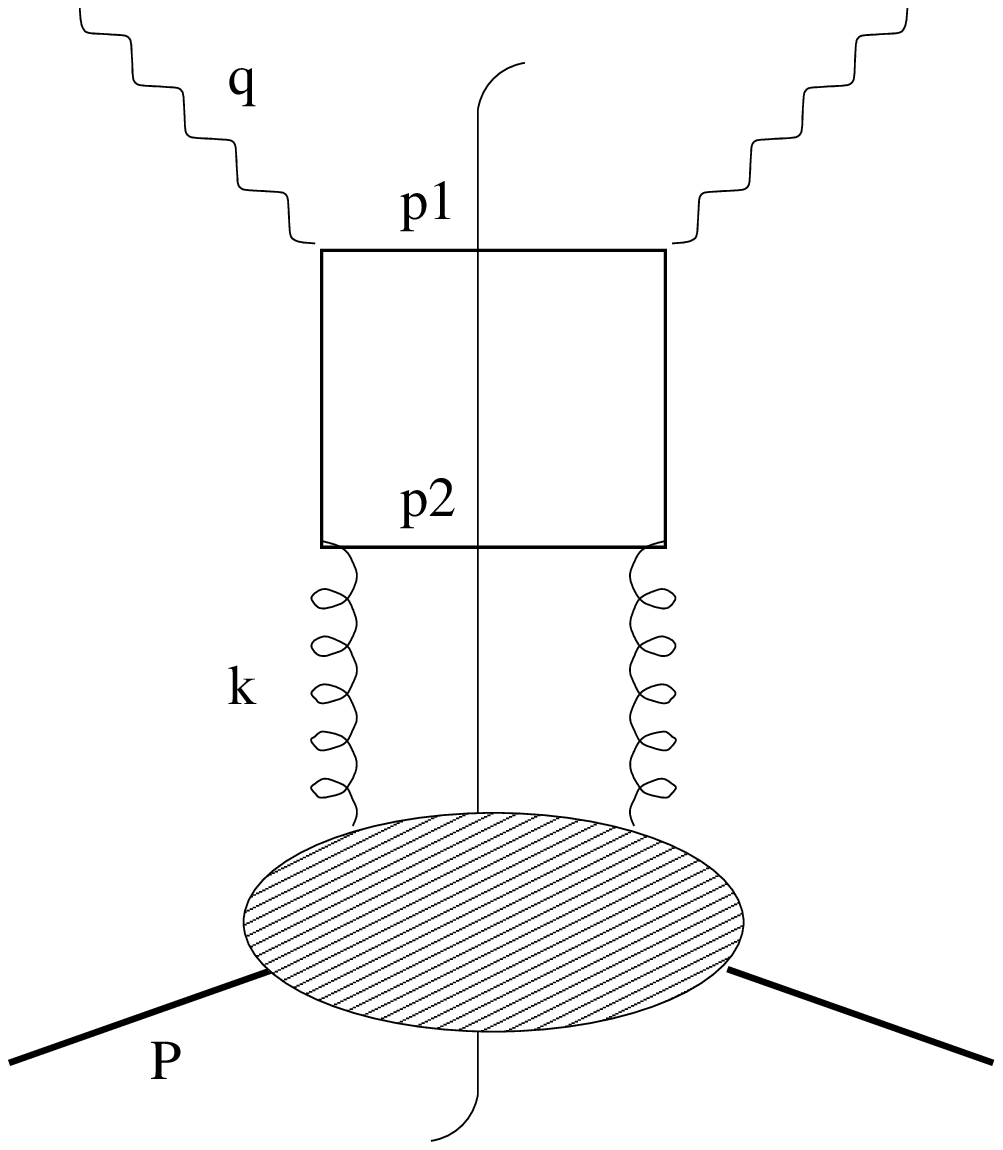}  }
  \;+\;
  \VC{\includegraphics[scale=0.5]{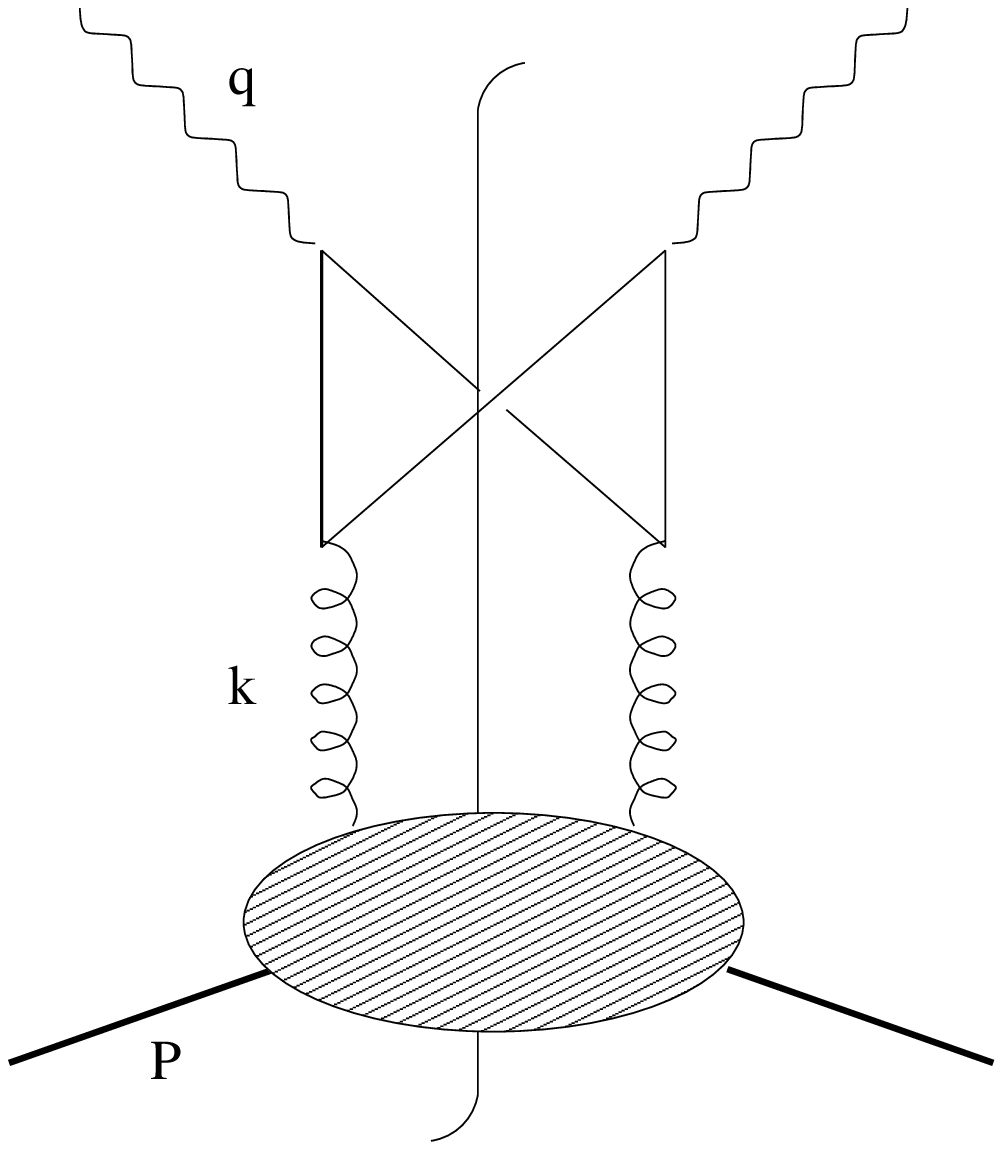}}
  $
  \caption{Photon-gluon fusion.}
  \label{fig:gamma.g.fusion}
\end{figure}

The problem that is addressed is nicely illustrated by considering
photoproduction of $c\bar{c}$ pairs.  In Figs.~\ref{fig:ccbar.1}, we
compare three methods of calculation carried out 
within the CASCADE event generator \cite{jung_salam_2000,*CASCADEMC}:
\begin{itemize}
\item Use of a conventional gluon density that is a function of parton
  $x$ alone.
\item Use of a $\kt$ density that is a function of parton $x$ and
  $\kt$.  These are the objects usually called ``unintegrated parton
  densities''.
\item Use of a doubly unintegrated density that is a function of parton
  $x$, $\kt$ and virtuality, that is, of the complete parton 4-momentum.
\end{itemize}
The partonic subprocess in all cases is the lowest order
photon-gluon-fusion process $\gamma+g\longrightarrow c+\bar{c}$ 
(Fig.~\ref{fig:gamma.g.fusion}).  Two differential cross sections are
plotted: one as a function of the transverse momentum of the $c\bar{c}$
pair, and the other as a function of the $x_\gamma$ of the pair.
By $x_\gamma$ is meant the fractional momentum of the photon carried by the
$c\bar{c}$ pair, calculated in the light-front sense as
\begin{displaymath}
 x_\gamma =  \frac{\sum_{i=c,\bar{c}} (E_i - p_{z\;i}) }{2 y E_e} 
    =  \frac{p_{c\bar{c}}^-}{q^-}.
\end{displaymath}
Here $E_e$ is the electron beam energy and the coordinates are
oriented so that the electron and proton beams are in the $-z$ and
$+z$ directions respectively.

In the normal parton model approximation for the hard scattering, the
gluon is assigned zero transverse momentum and virtuality, so that the
cross section is restricted to $\ptpair=0$ and $x_\gamma=1$, as
shown by the solid lines in Fig.\ \ref{fig:ccbar.1}(a,b).  When a
$\kt$ dependent gluon density is used, quite large gluonic $\kt$ can be
generated, so that the $\ptpair$ distribution is spread out in
a much more physical way, as given by the dashed line in Fig.\ 
\ref{fig:ccbar.1}(a).  But as shown in plot (b), $x_\gamma$ stays close to
unity. 
Neglecting the full recoil mass $m_{\rm rem}$ (produced in the shaded
subgraph in Fig~\ref{fig:gamma.g.fusion}) 
is equivalent of taking $k^2= -\kt^2/(1-x)$ with 
$k^2$ being the virtuality of the gluon in Fig.\ \ref{fig:gamma.g.fusion},
$\kt$ its transverse momentum and $x$ its light cone energy
fraction.  This gives a particular value to the gluon's $k^-$. 
When we also take into account the correct virtuality of  gluon, there
is no noticeable change in the $\ptpair$ distribution --- see Fig.\ 
\ref{fig:ccbar.1}(c) (dashed line) --- since that is already made broad by the
transverse momentum of the gluon.  But the gluon's $k^-$ is able to
spread out the $x_\gamma$ distribution, as in Fig.\ \ref{fig:ccbar.1}(d) with
the dashed line. This is equivalent with a proper treatment of the kinematics
and  results in   $k^2= - (\kt^2 + x m_{\rm rem}^2) / (1-x)$, where
$m_{\rm rem}$ is the invariant mass of the beam remnant, the part of
the final state in the shaded blob in Fig.\ \ref{fig:gamma.g.fusion}.
This change can be particularly significant if $x$ is not very small.

Note that if partons are assigned approximated 4-momenta during
generation of an event in a MC event generator, the momenta need to be
reassigned later, to produce an event that conserves total 4-momentum.
The prescription for the reassignment is somewhat arbitrary, and it is
far from obvious what constitutes a correct prescription, especially
when the partons are far from a collinear limit.  A treatment with
fully unintegrated pdfs should solve these problems.

\begin{figure}
  \centering
  \includegraphics[scale=0.6]{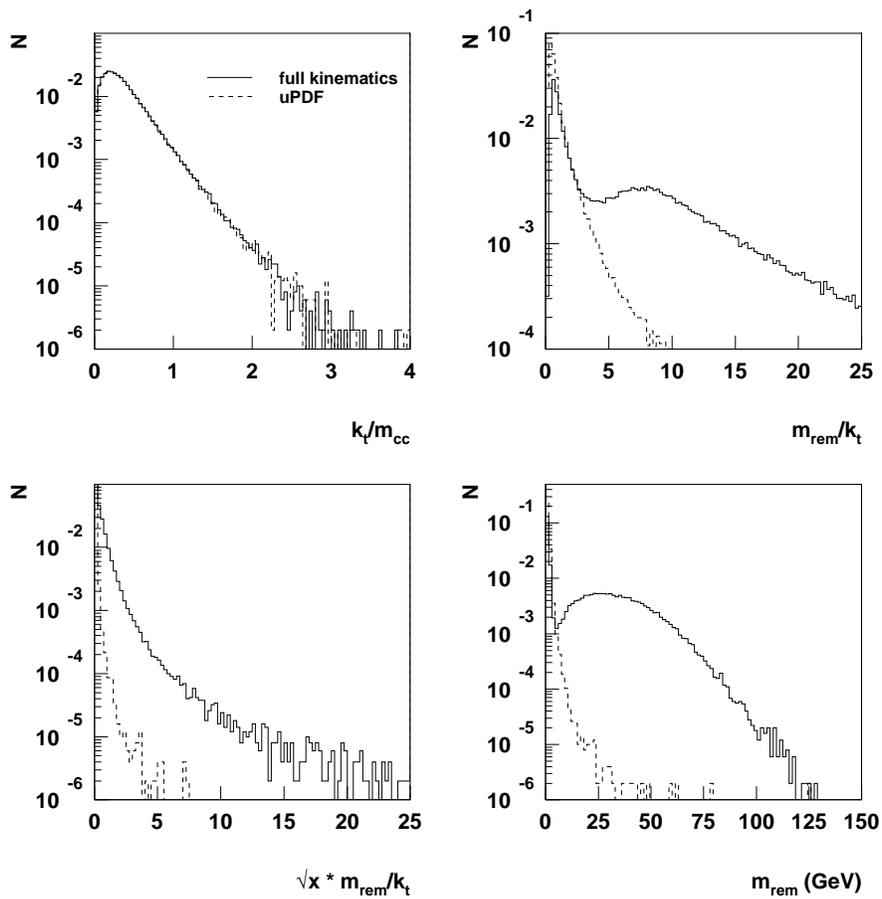}
  \caption{
    Comparison of distributions in partonic variables between
    calculations with full parton kinematics and with ordinary
    unintegrated pdfs.} 
  \label{fig:ccbar.2}
\end{figure}

If, as we claim, an incorrect treatment of parton kinematics changes
certain measurable cross sections by large amounts, then we should
verify directly that there are large discrepancies in the
distributions in partonic variables themselves.  We see this in 
Fig.\ \ref{fig:ccbar.2}.  Graph (a) plots the gluonic transverse
momentum divided by the charm-pair mass.  As is to be expected, the
typical values are less than one, but there is a long tail to high
values.  But the use of full parton kinematics does not have much of
an effect, the unintegrated parton distributions already providing
realistic distributions in transverse momentum.

On the other hand, a simple collinear approximation for showering sets
the remnant mass, $m_{\rm rem}$, to zero.  As can be seen from the
formulae for the gluon virtuality, this only provides a good
approximation to the gluon kinematics if $m_{\rm rem}$ is much less
than $\kt$.  In reality, as we see from graph (b), there is a long
tail to large values of $m_{\rm rem}/\kt$, and the tail is much bigger
when correct kinematics are used.  A more correct comparison uses
$xm_{\rm rem}^2$, with an extra factor of $x$.  Even then, there is a
large effect, shown in graph (c).  The vertical scale is logarithmic,
so the absolute numbers of events are relatively small, but the tail
is broad.  Finally, graph (d) shows that the distribution in $m_{\rm
  rem}$ itself is very broad, extending to many tens of GeV.  This
again supports the argument that unless a correct treatment of parton
kinematics is made, very incorrect results are easily obtained.

It is important to note that, for the cross sections themselves, the
kinematic variables used in Fig.\ \ref{fig:ccbar.1} are normal
ones that are in common use.  Many other examples are easily
constructed.  Clearly, the use of the simple parton-model kinematic
approximation gives unphysically narrow distributions.  The correct
physical situation is that the gluon surely has a distribution in
transverse momentum and virtuality, and for the considered cross
sections neglect of parton transverse momentum and virtuality leads to
wrong results.  It is clearly better to have a correct starting point
even at LO, for differential cross sections such as we have plotted.

\section{Kinematic approximations}

The standard treatment of parton kinematics involves replacing the
incoming parton momentum $k$ by its plus component only:
$k^\mu\mapsto\hat{k}^\mu\equiv(k^+,0,0_T)$.  There are actually two parts to this.
The first is to neglect the $^-$ and transverse components of $k$ with
respect to the large transverse momenta in the 
calculation
of the numerical value of the hard-scattering amplitude; 
this
is a legitimate approximation, readily corrected by higher order terms
in the hard scattering.  The second part is to change the kinematics
of the final-state particles, $p_1$ and $p_2$, so that their sum is $q$
plus the approximated gluon momentum.  It is this second part that is
problematic, for it amounts to the replacement of the momentum
conservation delta function $\delta^{(4)}(k+q-p_1-p_2)$ by
$\delta^{(4)}(\hat{k}+q-p_1-p_2)$.  These delta-functions are infinitely
different, point-by-point.  Only when integrated with a sufficiently
smooth test function can they be regarded as being approximately the
same, as in a fully inclusive cross section.

In an event generator, the effect is to break momentum conservation,
which is restored by an ad hoc correction of the parton kinematics.  
Note that the change of parton kinematics is only in the hard
scattering, i.e., in the upper parts of the graphs.  Parton kinematics 
are left unaltered within the parton density part, and the integrals
over $\kt$ and virtuality are part of the standard definition of
integrated pdfs.

The situation is ameliorated by inclusion of NLO terms, and perhaps
also by some kind of resummation.  But these do not correct the
initial errors in the approximation, and lead to a very restricted
sense in which the derivation of the cross section can be regarded as
valid.  Furthermore, when much of the effect of NLO terms is to
correct the kinematic approximations made in LO, this is an
inefficient use of the enormous time and effort going into NLO
calculations.  A case in point is the BFKL equation, where $70\%$ of
the (large) NLO corrections are accounted for \cite{BFKL.NLO} by
the correction of kinematic constraints in the LO calculation.

\section{Conclusions}

The physical reasoning for the absolute necessity of fully
unintegrated densities is, we believe, unquestionable. 
Therefore it is
highly desirable to reformulate perturbative QCD methods in terms of
doubly unintegrated parton densities from the beginning.  A full
implementation will be able to use the full power of calculations at
NLO and beyond.  

Among other things, a full implementation, as in \cite{CZ}, will
provide extra factorization formulae for obtaining the values of the
unintegrated densities at large parton transverse momentum and
virtuality.  This will incorporate all possible perturbatively
calculable information, so that the irreducible nonperturbative
information, that must be obtained from data, will be at low
transverse momentum and virtuality.  In addition, the implementation
will quantify the relations to conventional parton densities.  With
the most obvious definitions, the integrated pdfs are simple integrals
of the unintegrated densities.  However, in full QCD a number of
modifications are required \cite{CZ,What}, so that the relations
between integrated and unintegrated pdfs are distorted. 

The fact that we propose new and improved methods does not invalidate
old results in their domain of applicability.  The work of Watt,
Martin and Ryskin, and of Collins and Zu provides a start on this
project; but much remains to be done to provide a complete
implementation in QCD; for example, there is as yet no precise, valid,
and complete gauge-invariant operator definition of the doubly
unintegrated densities in a gauge theory.

The outcome of such a program should have the following results:
\begin{enumerate}
\item Lowest order calculations will give a kinematically much more
  realistic description of cross sections.  This may well lead to NLO
  and higher corrections being much smaller numerically than they
  typically are at present, since the LO description will be better.
  
\item It will also obviate the need for separate methods (resummation
  or the CSS technique), which are currently applied to certain
  individual cross sections like the transverse-momentum distribution
  for the Drell-Yan process.  All these and others will be subsumed
  and be given a unified treatment.  

\item A unified treatment will be possible for both inclusive cross
  sections using fixed order matrix element calculations and for
  Monte-Carlo event generators.
  
\item For a long-term theoretical perspective, the doubly unintegrated
  distributions will interface to methods of conventional quantum
  many-body physics much more easily than regular parton densities,
  whose definitions are tuned to their use in ultra-relativistic
  situations. 

\end{enumerate}
This program is, of course, technically highly nontrivial if it is to
be used in place of conventional methods with no loss of predictive
power.  A start is made in the cited work.

Among the main symptoms of the difficulties are that the most obvious
definition of a fully unintegrated density is a matrix element of two
parton fields at different space-time points, which is not
gauge-invariant.  It is often said that the solution is to use a
light-like axial gauge $A^+=0$.  However, in unintegrated densities,
this leads to divergences --- see \cite{What} for a review --- and the
definitions need important modification, in such a way that a valid
factorization theorem can be derived. 

We also have to ask to what extent factorization can remain true in a
generalized sense.  Hadron-hadron collisions pose a particular problem
here, because factorization needs a quite nontrivial cancellation
arising from a sum over final-state interactions.  This is not
compatible with simple factorization for the exclusive components of
the cross section, and makes a distinction between these processes and
exclusive components of DIS, for example.

\section*{Acknowledgments}

This work is supported in part by the U.S. DOE.

\providecommand{\etal}{et al.\xspace}
\providecommand{\coll}{Coll.}
\catcode`\@=11
\def\@bibitem#1{%
\ifmc@bstsupport
  \mc@iftail{#1}%
    {;\newline\ignorespaces}%
    {\ifmc@first\else.\fi\orig@bibitem{#1}}
  \mc@firstfalse
\else
  \mc@iftail{#1}%
    {\ignorespaces}%
    {\orig@bibitem{#1}}%
\fi}%
\catcode`\@=12

\end{document}